\documentclass[12pt]{JHEP3}

\usepackage{amsmath,amssymb,axodraw,graphicx}

\newcommand\fverb{\setbox\pippobox=\hbox\bgroup\verb}
\newcommand\fverbdo{\egroup\medskip\noindent%
                        \fbox{\unhbox\pippobox}\ }
\newcommand\fverbit{\egroup\item[\fbox{\unhbox\pippobox}]}
\newbox\pippobox

\title{\Large\bf Dynamical $U(1)_R$ Breaking in the Metastable Vacua}

\preprint{SNUTP 07-008}

\author{
Hae Young Cho  and Jong-Chul Park\\
  Department of Physics and Astronomy and Center for Theoretical Physics,
  Seoul National University, Seoul 151-747, Korea \\
  E-mail: \email{hycho@phya.snu.ac.kr,\ jcpark@phya.snu.ac.kr} }

\maketitle

\abstract{In the Intriligator-Seiberg-Shih model, we parametrize
spontaneous breaking of $U(1)_R$ symmetry with two gauge singlets
with R-charges 1 and --1. These singlets can play the role of the
messengers. The messenger scale is dynamically generated, and
hence there is no hierarchy problem between the supersymmetry
breaking scale and the messenger scale. In the gauge mediation
scenario, supersymmetry breaking scale turns out to be around
$\mathcal{O}(10^6)\textrm{\ GeV}$.}


\begin{document}

\def\lsl{ l \hspace{-0.45 em}/}
\def\ksl{ k \hspace{-0.45 em}/}
\def\qsl{ q \hspace{-0.45 em}/}
\def\psl{ p \hspace{-0.45 em}/}
\def\ppsl{ p' \hspace{-0.70 em}/}
\def\dsl{ \partial \hspace{-0.45 em}/}
\def\Dsl{ D \hspace{-0.55 em}/}
\def\matrix{ \left(\begin{array} \end{array} \right) }
\def\frsqotw{\frac{1}{\sqrt2}}
\def\frsqoth{\frac{1}{\sqrt3}}
\def\frsqtth{\sqrt{\frac{2}{3}}}
\def\frsqo{\frac{\omega}{\sqrt{3}}}
\def\frsqoo{\frac{\omega^2}{\sqrt{3}}}

\def\hf{\textstyle{\frac12~}}
\def\hff{\textstyle{\frac13~}}
\def\hfg{\textstyle{\frac23~}}
\def\DQ{$\Delta Q$}

\def\Qem{Q_{\rm em}}

\thispagestyle{empty}

\baselineskip 0.7cm

\section{Introduction}

Supersymmetry (SUSY) is still considered to be one of the best
candidates for the solution of the gauge hierarchy problem. It
explains the Higgs mass problem elegantly, but the question
$\lq\lq$Why is the electroweak scale so much smaller compared to the
Planck scale?" needs a judicious setup for SUSY breaking. About
thirty years ago Witten studied this question and suggested that the
discrepancy between two scales might be understood if SUSY is broken
dynamically \cite{Witten:1981kv}. Pre-2006 studies along this line
have not shown any compelling model applicable to the real world
phenomenology \cite{Affleck:1984xz, Dine:1993yw, Dine:1994vc,
Dine:1995ag, Poppitz:1996fw, Arkani-Hamed:1997jv}. In some of these
approaches there were SUSY preserving vacua by the interaction
between the messenger sector and the SUSY breaking sector. Thus,
these models already had the feature of metastable SUSY breaking
vacua \cite{Dine:1993yw, Dine:1994vc, Dine:1995ag}. The idea of
metastable SUSY breaking false vacua with SUSY preserving true
minimum only in the SUSY breaking sector was introduced about ten
years ago \cite{Dimopoulos:1997ww, Luty:1997ny, Dimopoulos:1997je}.
Recently, Intriligator, Seiberg, and Shih (ISS) gave an explanation
that the metastability of the supersymmetry breaking vacua could be
generic in SUSY gauge theories. Therefore, this argument could
simplify and enrich the realm of the model building possibilities
\cite{Intriligator:2006dd}. This work has especially attractive
features in that it can be accommodated in string theory
\cite{Ooguri:2006bg, Franco:2006ht, Kim:2007zj} and one can obtain
low energy supersymmetry breaking.

Independently of these arguments on SUSY breaking, $U(1)_R$
symmetry breaking showed an important role on SUSY breaking
itself. The origin of $U(1)_R$ symmetry is related to the rotation
in the superspace which is defined by the fermionic coordinates,
\begin{equation}
U(1)_R: \theta \rightarrow e^{-i\alpha}\theta.
\end{equation}
If we consider gravitational effect, $U(1)_R$ symmetry is explicitly
broken by gravitational effect. We don't, however, consider this
case here. Early in the 1990s Nelson and Seiberg showed that
$U(1)_R$ symmetry is necessary for breaking SUSY spontaneously in
models with generic and calculable superpotentials
\cite{Nelson:1993nf}. Therefore, the effect of $U(1)_R$ symmetry is
protecting the SUSY  breaking minimum from being invaded by the SUSY
preserving minima. This pattern is also applicable to the case of
the meta-stable vacua. Instead of the exact $U(1)_R$ symmetry, in
the meta-stable vacuum case there exists an accidental and
approximate one near the origin in the field configuration space.
The effect of this symmetry is again to protect the meta-stable
vacuum. In any case, it is necessary to have the $U(1)_R$ symmetry
to obtain a stable SUSY breaking global or local minimum.
Nevertheless, the $U(1)_R$ symmetry should be broken to obtain the
soft gaugino masses and to make the R-axion heavy \cite{Dine:1993yw,
Bagger:1994hh, Luty:1997ny}.

Related to the ISS model, many studies have been performed to
break the accidental $U(1)_R$ symmetry. Some studies were done in
direct mediation setup with an explicit R-symmetry breaking term
\cite{Kitano:2006xg,Csaki:2006wi}. There have been studies on
spontaneous R-symmetry breaking by introducing $U(1)$ gauge
interaction in the direct gauge mediation scheme
\cite{Csaki:2006wi}, but these studies have a problem with the
Landau pole. The direct mediation scheme in the ISS scheme has
problems because the ISS model has too many massive flavors. Some
methods employed to avoid the Landau pole problem lead to too high
SUSY breaking scale \cite{Kitano:2006xg}. On the other hand, the
Landau pole is inescapable if the SUSY is broken at the low scale
\cite{Csaki:2006wi}. Therefore, the ordinary gauge mediation is
more persuasive in this sense. Some studies related to the
R-symmetry breaking in the ISS setup introduce an explicit
breaking term, which comes from the interaction between the
messengers and the Goldstino super multiplet in the ordinary gauge
mediation \cite{Murayama:2006yf,Aharony:2006my}, but the messenger
masses are introduced by hand. Thus, the hierarchy between the
SUSY breaking scale and the messenger scale remains unsolved.

In this paper, we study dynamical spontaneous $U(1)_R$ symmetry
breaking in a modified ISS set up without the help of any explicit
$U(1)_R$ breaking terms. We introduce two singlets with R-charges 1
and --1, respectively, to keep the superpotential invariant under
the $U(1)_R$ symmetry. By doing this, we can avoid the two
superficial dilemma: obtaining $U(1)_R$ symmetry and breaking it. We
find the $U(1)_R$  and SUSY breaking meta-stable minima without
fine-tuning. Since the original work of ISS is related to the SQCD,
we do not want to harm the good property that SUSY is restored at
high energy scale by the dynamical interaction
\cite{Intriligator:2006dd}. We find that our vacua are near the
origin in the field space. It means that we can keep the strong
points of the ISS model. The vacua can survive the transition to the
true SUSY preserving vacua as long as its lifetime which is much
longer than the age of our universe.  Through this study, we obtain
the dynamically generated messenger scale. At the same time, this
scale is the SUSY breaking scale, and hence we can build a realistic
model without introducing the messenger scale by hand. In addition,
the new singlet fields can be used as the messenger fields in the
gauge mediation scenario.

In Sec. 2, we  briefly review the ISS model, and  consider the
$U(1)_R$ symmetry breaking in the O'Raifeartaigh type SUSY
breaking. In Sec. 3, we introduce the model. In Sec. 4,
phenomenological implications are commented.

\section{The ISS model and $U(1)_R$ symmetry}

The O'Raifeartaigh model is the simplest mechanism breaking SUSY
spontaneously \cite{O'Raifeartaigh:1975pr}. The basic setup of the
ISS model is related to this mechanism. In the early 1990s, the
holomorphic property of SUSY gauge theory together with the global
symmetry properties showed the good vacuum  properties, such as
the moduli space structure and the duality between the electric
theory and its magnetic dual theory, which ordinary gauge theories
do not possess \cite{Seiberg:1994bz, Seiberg:1994pq,
Intriligator:1995au}.  SUSY QCD (SQCD) was classified by the
number of flavors $N_f$ and the number of color $N_c$. ISS started
with the magnetic dual gauge theory which is infrared free, for
which the electric theory is asymptotically free. This happens for
$N_c<N_f<\frac{3}{2}N_c$.

The field contents in the magnetic dual theory are,
\begin{equation}
\begin{array}{c|ccc}
 & \Phi  & \varphi & \tilde \varphi \\ \hline
 SU(N) & 1 & \square & \overline{\square}  \\
SU(N_f) & {\rm adj}+1 & \square & \overline{\square}  \\
U(1)_R & 2 & 0 & 0  \\
\end{array} ,
\end{equation}
where $N=N_f-N_c$. Then the superpotential consistent with the
symmetry is given by
\begin{equation}
W = h{\rm Tr}\tilde \varphi \Phi \varphi - h\mu^2{\rm Tr}\Phi.
\end{equation}
If we assume that there is a kind of strongly coupled interaction,
then we can make $\mu$ small enough by retrofitting as
\begin{equation}
\mu=\frac{\Lambda_s^3}{M_P^2},
\end{equation}
where $\Lambda_s$ is the confining scale of a strong dynamics
\cite{Dine:2006gm}. Then SUSY is broken via the incompatibility
among the F-flat conditions, i.e. by the rank condition, at the
origin of the field space. The tree level potential is
\begin{equation}
V_{cl}=(N_f-N)|h^2\mu^4|
\end{equation}
with the flat direction given by
\begin{equation}
\Phi = \left(%
\begin{array}{cc}
  0 & 0 \\
  0 & X \\
\end{array}%
\right), \ \
\varphi = \left(%
\begin{array}{cc}
  \varphi_0 \\
  0 \\
\end{array}%
\right),\ \
\tilde \varphi_0 ^T = \left(%
\begin{array}{cc}
  \tilde \varphi_0 \\
  0 \\
\end{array}%
\right)\\
\end{equation}\\
with $\tilde \varphi_0 ^T \varphi_0 = \mu^2 {\bf1}_N$.

If we turn on the $SU(N)$ which is IR free, there exists a scale
$\Lambda_m$, above which the theory turns to be strongly coupled
theory. The holomorphic gauge coupling of $SU(N)$ is given by
\begin{equation}
e^{-8\pi^2/g^2(E)+i\theta}=\left(\frac{E}{\Lambda_m}\right)^{N_f-3N_c}.
\end{equation}
Now we consider quantum correction to the potential, then the
theory turns out to maintain the supersymmetry breaking feature at
the origin even after we introduce the gauge interaction. However,
the gauge interaction plays a crucial role in the other place of
the field configuration space. By non-perturbative effect we get
the effective potential as
\begin{equation}
W_{low}=N(h^{N_f}\Lambda_{m}^{-(N_f-3N)}\det\Phi)^{1/N}-h\mu^2Tr\Phi.\label{dynW}
\end{equation}
Next, by investigating the F-flat directions, we obtain a result
that there exist supersymmetry preserving vacua, i.e. all the F
terms vanish, with the value of
\begin{equation}
\langle h\Phi
\rangle=\Lambda_m\epsilon^{2N/(N_f-N)}{\bf1}_{N_f}=
\mu\frac{1}{\epsilon^{(N_f-3N)/(N_f-N)}}{\bf1}_{N_f},
\ \ \textrm{where} \ \ \epsilon \equiv \frac{\mu}{\Lambda_m}.
\end{equation}
The longevity of the metastable vacua is guaranteed by this
inequality for $\epsilon \ll 1,$
\begin{equation}
|\mu| \ll | \langle h\Phi \rangle| \ll |\Lambda_m|.
\end{equation}
If we return to the origin where the gauge interaction can be
ignored, we find that there exists an accidental R-symmetry
induced by quantum corrections. The existence of supersymmetry
preserving vacua means that there is no exact $U(1)_R$ symmetry.

Nelson and Seiberg showed that generic and calculable models need
$U(1)_R$ symmetry to break SUSY spontaneously \cite{Nelson:1993nf}.
However, SUSY is not spontaneously broken in spite of $U(1)_R$ when
R-charges of all matter fields are either 2 or 0. This implies that
R-charges of field contents are closely related to the SUSY breaking
as mentioned above. Recently, Shih gave an explanation for the
R-symmetry breaking, which is just  a necessary condition for
$U(1)_R$ symmetry breaking minimum in O'Raifeartaigh type models.
Including the radiative corrections as in the Coleman-Weinberg
potential, the pseudo-moduli gain masses at the origin in generic
cases. He found that introducing some fields with R-charges except 0
and 2 can make the masses of the moduli negative by quantum
corrections, but this depends on the condition of the parameter
space. The negative masses of moduli mean that $U(1)_R$ symmetry is
spontaneously broken via quantum corrections. Therefore, he led the
conclusion that \emph{in O'Raifeartaigh type SUSY breaking models,
it is necessary to have matter fields of which R-charges are given
neither 2 nor 0 for spontaneous $U(1)_R$ breaking}
\cite{Shih:2007av}.

\section{Spontaneous breaking of $U(1)_R$ in a modified ISS  model}

\subsection{Model}

In the O'Raifeartaigh-type models where all fields have R-charges of
0 or 2 only, the $U(1)_R$ symmetry is not spontaneously broken.
Therefore, in order for $U(1)_R$ to be spontaneously broken, there
has to be at least one field in the model with R-charge different
from 0 and 2. \cite{Shih:2007av}

Thus, we introduce new fields, $A$ and $B$, with R-charge different
from 0 and 2 to the original ISS type model. The matter fields we
introduce are for $N_f > N$ :
\begin{equation}
\begin{array}{c|ccccc}
 & \Phi  & \varphi & \tilde \varphi & A & B\\ \hline
 SU(N) & 1 & \square & \overline{\square} &  1 & 1 \\
SU(N_f) & {\rm adj}+1 & \square & \overline{\square} &  1 & 1 \\
U(1)_R & 2 & 0 & 0 & 1 & -1 \\
\end{array}  .
\end{equation}
Taking the canonical K\"ahler potential, the generic tree-level
superpotential is
\begin{equation}
W = h{\rm Tr}\tilde \varphi \Phi \varphi - h\mu^2{\rm Tr}\Phi +
\lambda AB{\rm Tr}\Phi + mA^2
\end{equation}
which respects the $U(1)_R$ symmetry. The classical moduli space is
obtained from
\begin{equation}
\begin{split}
\frac{\partial W}{\partial A} = \lambda B {\rm Tr} \Phi + 2mA,\ \
&\frac{\partial W}{\partial B} = \lambda A {\rm Tr} \Phi,\ \
\frac{\partial W}{\partial \Phi_{ij}} = h \tilde \varphi^i
\varphi^j
+ (\lambda AB - h \mu^2)\delta^{ij},\\
&\frac{\partial W}{\partial \phi} = h \tilde \varphi \Phi,\ \
\frac{\partial W}{\partial \tilde \phi} = h \Phi \varphi .
\end{split}
\end{equation}
The  vacua along the classical moduli space are
\begin{equation}
A=0,\ \ B= 0,\ \
\Phi = \left(%
\begin{array}{cc}
  0 & 0 \\
  0 & X \\
\end{array}%
\right), \ \
\varphi = \left(%
\begin{array}{cc}
  \varphi_0 \\
  0 \\
\end{array}%
\right),\ \
\tilde \varphi_0 ^T = \left(%
\begin{array}{cc}
  \tilde \varphi_0 \\
  0 \\
\end{array}%
\right)\label{pseudomoduli}
\end{equation}
 or
 \begin{align}
 A=0,\ \ B= {\rm arbitrary},\ \
&\Phi = \left(%
\begin{array}{cc}
  0 & 0 \\
  0 & X \\
\end{array}%
\right), \ \
\varphi = \left(%
\begin{array}{cc}
  \varphi_0 \\
  0 \\
\end{array}%
\right),\ \
\tilde \varphi_0 ^T = \left(%
\begin{array}{cc}
  \tilde \varphi_0 \\
  0 \\
\end{array}%
\right)\ \ {\rm where}\ \ {\rm Tr}X=0
\end{align}
with $\tilde \varphi_0 ^T \varphi_0 = \mu^2 {\bf1}_N$. Then, the
classical scalar potential is given by
\begin{equation}
V = (N_f-N) \left|h \mu^2 \right|^2.
\end{equation}

This model does not have a SUSY ground state because it is
impossible for $F_A$, $F_B$, and $F_{\Phi}$ terms to vanish
simultaneously. However, at
\begin{equation}
A=\frac{h\mu^2}{\lambda} \frac{1}{B},\ \  {\rm Tr}\Phi = 0,\ \
\varphi = 0,\ \  \tilde \varphi = 0 ,\label{runaway}
\end{equation}
the classical potential has a runaway direction toward $B
\rightarrow \infty$,
\begin{equation}
V = 4\left|mA \right|^2 = 4\left| \frac{h\mu^2m}{\lambda B}
\right|^2 \rightarrow 0 .
\end{equation}
No static vacuum exists along this direction, and SUSY is
asymptotically restored as $B \rightarrow \infty$.

\subsection{One-loop lifting of pseudo-moduli}

The minima of the tree-level scalar potential considered here
occur along the pseudo-moduli space (\ref{pseudomoduli}).
Expanding around the above classical moduli space
(\ref{pseudomoduli}),\footnote{We choose the case of $B=0$ and
$X=$ arbitrary even though the case of $B=$ arbitrary and Tr $X=0$
is also possible. If we take A and B as the messengers, we don't
have to consider the latter.}
\begin{equation}
\begin{split}
\Phi = \left(%
\begin{array}{cc}
  \delta \Phi_{11} & \delta \Phi_{12} \\
  \delta \Phi_{21} & X+\delta \Phi_{22} \\
\end{array}%
\right),\ \
& \varphi = \left(%
\begin{array}{cc}
  \mu Y+\delta \varphi_1 \\
  \delta \varphi_2 \\
\end{array}%
\right),\ \
\tilde \varphi_0 ^T = \left(%
\begin{array}{cc}
  \mu/Y+\delta \tilde \varphi_1 \\
  \delta \tilde \varphi_2 \\
\end{array}%
\right)\\
& A=\delta A,\ \ B=\delta B ,
\end{split}
\end{equation}
the superpotential can be expressed as
\begin{equation}
\begin{split}
W =& h{\rm Tr}\tilde \varphi \Phi \varphi - h\mu^2{\rm Tr}\Phi +
\lambda AB{\rm Tr}\Phi + mA^2\\
=& -h\mu^2{\rm Tr}(X+\delta\Phi_{22})+m(\delta A)^2\\
&+h{\rm Tr}[\delta\Phi_{11}(\mu
Y\delta\tilde\varphi_1+\delta\varphi_1\frac{\mu}{Y})
+\frac{\mu}{Y}\delta\Phi_{12}\delta\varphi_2
+\delta\tilde\varphi_2\delta\Phi_{21}\mu Y]\\
&+h{\rm Tr}[\delta\tilde\varphi_1\delta\Phi_{11}\delta\varphi_1
+\delta\tilde\varphi_1\delta\Phi_{12}\delta\varphi_2
+\delta\tilde\varphi_2\delta\Phi_{21}\delta\varphi_1
+(X+\delta\Phi_{22})\delta\varphi_2\delta\tilde\varphi_2]\\
&+\lambda\delta A\delta B{\rm
Tr}[\delta\Phi_{11}+(X+\delta\Phi_{22})] .
\end{split}\label{W-expand}
\end{equation}

Now let us consider the Coleman-Weinberg potential for the
pseudo-moduli({\it i.e.} $X$)\cite{Coleman:1973jx}
\begin{equation}
\begin{split}
V^{(1)}_{eff} = & \frac{1}{64\pi^2}{\rm STr} \left(
\mathcal{M}^4\log
 \frac{\mathcal{M}^2}{\Lambda^2}\right)\\
 \equiv & \frac{1}{64\pi ^2}\left[ {\rm Tr} \left( \mathcal{M}_B^4 \log
 \frac{\mathcal{M}_B^2}{\Lambda ^2} \right)
 -{\rm Tr} \left( \mathcal{M}_F^4 \log \frac{\mathcal{M}_F^2}{\Lambda ^2}\right)
 \right],
\end{split}\label{eff-pot}
\end{equation}
where $\mathcal{M}_B^2$ and $\mathcal{M}_F^2$ are the tree-level
boson and fermion mass matrices
\begin{equation}
\mathcal{M}_B^2 = \left(%
\begin{array}{cc}
  W^{\dagger}_{ik}W^{kj} & W^{\dagger}_{ijk}W^k \\
  W^{ijk}W^{\dagger}_k & W^{ik}W^{\dagger}_{kj} \\
\end{array}%
\right),\ \
\mathcal{M}_F^2 = \left(%
\begin{array}{cc}
  W^{\dagger}_{ik}W^{kj} & 0 \\
  0 & W^{ik}W^{\dagger}_{kj} \\
\end{array}
\right).
\end{equation}
As usual, $W^i$ stands for $\partial W/\partial \delta \varphi_i$.
From now on, we  work at  $Y=1$.

Before going further, we have to check whether there exist tachyonic
modes at the pseudo-moduli space of (\ref{pseudomoduli}). Among the
eigenvalues of $\mathcal{M}_B^2$, there are two modes which could be
potentially dangerous,
\begin{equation}
\begin{split}
&\lambda^2X^2+2m^2-\sqrt{4m^2\lambda
^2X^2+4hm\lambda^2\mu^2X+h^2\lambda^2\mu^4+4m^4}\\
&\lambda^2X^2+2m^2-\sqrt{4m^2\lambda
^2X^2-4hm\lambda^2\mu^2X+h^2\lambda^2\mu^4+4m^4} .
\end{split}\label{tachyons}
\end{equation}
There are no tachyonic modes for the following range of the
pseudo-modulus $X$:
\begin{equation}
|X| \gtrsim {\rm max}\{\mu, (4m\mu^2)^{1/3} \}\label{no-tachyonic}
\end{equation}
where we set the couplings $h\simeq\lambda\simeq1$. The two modes
become tachyonic outside of the range of (\ref{no-tachyonic}) and
the pseudo-moduli space (\ref{pseudomoduli}) is locally unstable
there. As a result, the fields can roll down to the supersymmetric
runaway vacua along the tachyonic directions
\cite{Intriligator:2007py, Ferretti:2007ec}.

\begin{figure}[t]
\begin{center}
\includegraphics[width=10cm]{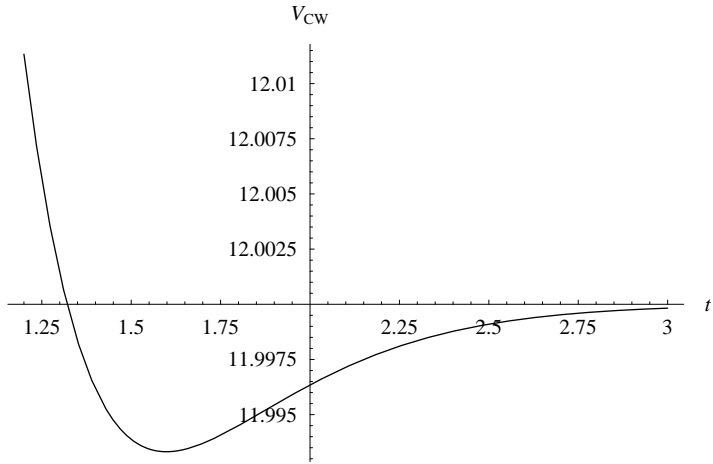}
\end{center}
\caption{Plot of the Coleman-Weinberg  potential without gauge
interaction terms. We neglect the numerical factor of
$\frac{1}{64\pi^2}$. We set $X/\mu=e^t$. } \label{fig:CW1}
\end{figure}

In the range of (\ref{no-tachyonic}) for the pseudo-modulus $X$,
there does not exist a tachyonic mode, and hence we can compute
the one-loop Coleman-Weinberg potential (\ref{eff-pot}) in that
pseudo-moduli space (\ref{pseudomoduli}). In order for the
$U(1)_R$ symmetry to be broken spontaneously, we need to find a
field, which has a non-zero VEV but charged with $U(1)_R$. In
fact, $X$ is a flat direction in the tree level and has a R-charge
2, but it is lifted and has a locally stable minimum at $X \sim
\mathcal{O}(1)\mu$ by the one-loop Coleman-Weinberg potential. The
$U(1)_R$ symmetry is spontaneously broken there. Note that
\begin{equation}
|\langle X\rangle| \sim \mathcal{O}(1)|\mu|\ \  \longrightarrow\ \
|\langle X\rangle| \ll |\Lambda_m| .
\end{equation}
Thus, our addition of new fields, $A$ and $B$, does not ruin the
merit of the original ISS model i. e. the longevity of the
metastable vacua. It is important that there exist stable minima
for a wide range of parameters of $h$, $\lambda$ and $m$:
\begin{equation}
h\gtrsim\mathcal{O}(0.1),\ \ \lambda\gtrsim\mathcal{O}(0.1),\ \
m\lesssim 1.3\mu .
\end{equation}
As an example, we show Fig. \ref{fig:CW1} for $h=1$, $\lambda=1$,
and $m=0.1\mu$.

\begin{figure}[t]
\begin{center}
\includegraphics[width=10cm]{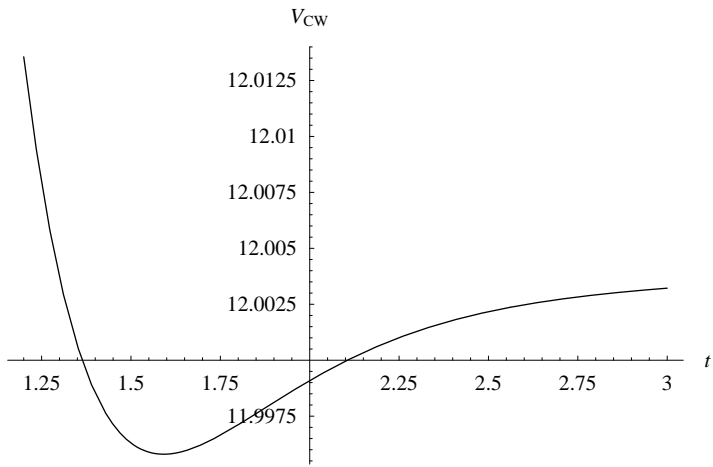}
\end{center}
\caption{Same as Fig. 1 except for the gauge interaction $g=0.1$.}
\label{fig:CW2}
\end{figure}

So far, we considered the superpotential without the gauge
interaction terms in the ISS model. Therefore, we also have to
check whether the gauge interaction terms spoil the basic feature
of our ISS model and whether the spontaneous $U(1)_R$ breaking
feature is maintained. It is easy to obtain the one-loop
Coleman-Weinberg potential in the presence of the gauge bosons and
gauginos, in the same range of the pseudo-modulus $X$ given above.
Obviously, if the gauge coupling $g$ is small, our result is not
modified significantly. As we can see in Fig. ~\ref{fig:CW2},
there exist stable minima at the order of $\mathcal{O}(1)\mu$ even
for the case of a bit large coupling. In addition, this effect
increases the slope of the Coleman-Weinberg potential at large
field values compared to the case without the gauge interaction
terms. As a result, the addition of the gauge interaction terms
rather improves the stability of the $U(1)_R$ breaking minima. A
plot of the Coleman-Weinberg potential with the gauge interaction
terms is given in Fig. \ref{fig:CW2} for $h=1$, $\lambda=1,
m=0.1\mu$, and $g=0.1$.

\section{Applications}

The exact $U(1)_R$ symmetry is broken at high energy scale, but it
seems that the approximate $U(1)_R$ can forbid the gaugino mass
terms. In our case, however, it is broken even at the low energy
scale. Therefore, we can generate the Majonara mass terms for the
gauginos. Since we obtain $U(1)_R$ symmetry breaking meta-stable
vacua dynamically and spontaneously,  the gaugino masses are
generated without introducing messenger mass by hand. In addition to
this, the messengers are not tachyonic, for it is always guaranteed
that $\langle F \rangle \leq \langle X \rangle ^2$ in the allowed
parameter range. If we accept the gauge mediation scheme, then the
messenger mass scale will be determined by the scales of
$\langle\textrm{Tr} \Phi\rangle =\langle X \rangle > \mu$ and
$\langle F \rangle \sim h\mu^2$. In the gauge mediation scheme
\footnote{For the review, see \cite{Giudice:1998bp}.}, we get the
gaugino soft mass terms as
\begin{equation}
\begin{split}
m_{\frac{1}{2}}(t)=&k_{r}\frac{\alpha_{r}(t)}{4\pi}\Lambda_G\\
\Lambda_G=&\sum_{i=1}^{N_{f}}n_i\frac{F_i}{M_i}g(F_i/M_i^2)\\
g(x)=&\frac{1}{x^2}\left[(1+x)\ln{(1+x)}+(1-x)\ln{(1-x)}\right],
\end{split}
\end{equation}
where $\alpha_{r}(t)$ is related to the visible sector gauge
coupling at the messenger scale. $k_{r}$ and $n_{i}$ are constants
of $\mathcal{O}(1)$, which depend on the messenger structure
\cite{Martin:1996zb}. A rough estimation of
$m_{\frac{1}{2}}\sim\mathcal{O}(1) \textrm{TeV}$ gives a
phenomenologically acceptable parameter range for $h$, $\lambda$ and
$m$. Namely, we get $\mu \sim \mathcal{O}(10^6)\textrm{GeV}$. Even
though we introduced the messenger fields, it is found that the
structure of the potential is not dangerous. As we have seen in Sec
3.2, the mass scales of the newly introduced singlets are determined
by the requirement of $U(1)_R$ symmetry breaking stable minima.
Moreover, the newly introduced singlets have an interaction term
with the Goldstino supermultiplet fields (i.e. X), and so they can
be used as the messenger fields. However, we do not consider any
specific models here. Their phenomenological applications are left
for future communication.

Now we will turn to the cosmological aspect. In the original ISS,
$U(1)_R$ symmetry is broken in the microscopic view point. On the
other hand, in the macroscopic view point R-symmetry is broken by
the non-perturbative term in (\ref{dynW}). Thus, this R-symmetry
is not exact but approximate and accidental. A spontaneously
broken approximate global symmetry gives arise to a pseudo
goldstone boson. In our model, $U(1)_R$ is an approximate symmetry
because of the $\Lambda_m$ suppressed term and what we have
obtained are the vacua which break it spontaneously. Therefore, we
do not have to worry about the R-axion.

\section{Conclusion}

We have studied the radiatively generated spontaneous $U(1)_R$
symmetry breaking in the ISS setup. We introduced two gauge singlet
fields with R-charges 1 and -1 respectively to keep the
superpotential generic up to $U(1)_R$ symmetry. Since the $U(1)_R$
is an approximate symmetry, we need not worry about the R-axion. For
the study of the pseudo-moduli space, we have found the
spontaneously broken radiatively generated meta-stable vacua with a
large range of parameter space. We can easily satisfy the
meta-stability of our vacua because it appears to be at the scale of
O($\mu$) where $\mu$ can be made small by the retrofitting argument.
We can obtain the dynamically generated messenger scale which turns
out to be the SUSY breaking scale. As a result, we can build a
realistic model without introducing the messenger scale by hand. In
addition, new singlets can act as the messengers. Finally, if the
gauge mediation scheme is used for the SUSY breaking, the SUSY
breaking scale is $\mathcal{O}(10^6)\textrm{GeV}$.

\begin{acknowledgments}
We thank Jihn E. Kim, H. D. Kim, I. W. Kim, and Sungjay Lee for
useful discussions. This work was supported in part by the Korea
Research Foundation Grant funded by the Korean Goverment(MOEHRD)
(KRF-2005-084-C00001). HC is also supported by the BK21 program of
Ministry of Education, and by the Center for Quantum Spacetime
(CQUeST),  Sogang University (the KOSEF grant R11-2005-021).
\end{acknowledgments}


\end{document}